\documentclass[11pt]{article}

\usepackage{dcolumn}
\usepackage{amsmath}
\usepackage{amssymb}
\usepackage{graphicx}
\usepackage{calrsfs}
\usepackage{bm}
\usepackage[table,xcdraw]{xcolor}
\usepackage{enumitem}
\usepackage{hyperref}

\newcommand{\Ord}{\mathrm{O}}

\newcommand{\argmax}{\mathop{\textrm{argmax}}\limits}

\newcommand\etal{\textit{et~al.}}

\newcommand{\review}[1]{{\color{black} #1}}

\begin{document}

\title{Normalized mutual information is a biased measure
for classification and community detection}

\author{Maximilian Jerdee$^{1,2}$, Alec Kirkley$^{3,4,5}$, and
Mark Newman$^{1,6,*}$\\[2ex]
$^1$Department of Physics, University of Michigan, Ann Arbor, Michigan 48109, USA\\
$^2$Santa Fe Institute, Santa Fe, New Mexico 87501, USA\\
$^3$Institute of Data Science, University of Hong Kong, Hong Kong\\
$^4$Department of Urban Planning and Design, University of Hong Kong, Hong Kong\\
$^5$Urban Systems Institute, University of Hong Kong, Hong Kong\\
$^6$Center for the Study of Complex Systems, University of Michigan,\\
Ann Arbor, Michigan 48109, USA\\[1ex]
$^*$Corresponding author, email mejn@umich.edu}

\maketitle

\begin{abstract}
Normalized mutual information is widely used as a similarity measure for evaluating the performance of clustering and classification algorithms.  In this paper, we argue that results returned by the normalized mutual information are biased for two reasons: first, because they ignore the information content of the contingency table and, second, because their symmetric normalization introduces spurious dependence on algorithm output.  We introduce a modified version of the mutual information that remedies both of these shortcomings.  As a practical demonstration of the importance of using an unbiased measure, we perform extensive numerical tests on a basket of popular algorithms for network community detection and show that one's conclusions about which algorithm is best are significantly affected by the biases in the traditional mutual information.
\end{abstract}

\section{Introduction}
\label{sec:intro}

A common task in data analysis is the comparison of two different labelings of a set of objects.  How well do demographics predict political affiliation?  How accurately do blood tests predict clinical outcomes?  These are questions about how similar one labeling of individuals/items/results is to another and they are commonly tackled using the information theoretic measure known as mutual information, in which an experimental or computational measure of some kind is compared against a ground truth.

Mutual information works by asking of how efficiently we can describe one labeling if we already know the other~\cite{CT06}.  Specifically, it measures how much less information it takes to communicate the first labeling if we know the second versus if we do not.  As an example, mutual information is commonly used in network science to evaluate the performance of algorithms for network community detection~\cite{DDDA05}.  One takes a network whose community structure is already known and applies a community detection algorithm to it to infer the communities.  Then one uses mutual information to compare the output of the algorithm to the known correct communities.  Algorithms that consistently achieve high mutual information scores are considered good.  We will use this application as an illustrative example later in the paper.

Mutual information has a number of appealing properties as a tool for comparing labelings.  It is invariant under permutations of the labels, so that, for example, two divisions of a network into communities that differ only in the numbering of the communities will be correctly identified as being the same.  It also returns sensible results in cases where the number of distinct label values is not the same in the two labelings.  However, the standard mutual information measure also has some significant shortcomings, and two in particular that we highlight in this paper.  First, it has a bias towards labelings with too many distinct label values.  For instance, a community detection algorithm that incorrectly divides networks into significantly more groups than are present in the ground truth can nonetheless achieve a high mutual information score.  A number of approaches for correcting this flaw have been proposed.  One can apply direct penalties for incorrect numbers of groups~\cite{AP17} or subtract correction terms based on the average mutual information over some ensemble of candidate labelings~\cite{VEB10,GA17} or on the statistics of the contingency table~\cite{NCY20,JKN24}.  For reasons we discuss shortly, we favor the latter approach, which leads to the measure known as the reduced mutual information.

The second drawback of the mutual information arises when the measure is normalized, as it commonly is to improve interpretability.  The most popular normalization scheme creates a measure that runs between zero and one by dividing the mutual information by the arithmetic mean of the entropies of the two labelings being compared~\cite{FJ03}, although one can also normalize by the minimum, maximum, or geometric mean of the entropies.  As we demonstrate in this paper, however, these normalizations introduce biases into the results by comparison with the unnormalized measure, because the normalization factor depends on the candidate labeling as well as the ground truth.  This effect can be large enough to change scientific outcomes, and we provide examples of this phenomenon.

In order to avoid this latter bias, while still retaining the interpretability of a normalized mutual information measure, we favor normalizing by the entropy of the ground-truth labeling alone.  This removes the source of bias but introduces an asymmetry in the normalization.  At first sight this asymmetry may seem undesirable, and previous authors have gone to some lengths to avoid it.  Here, however, we argue that it is not only justified but actually desirable, for several reasons.  First, many of the classification problems we consider are unaffected by the asymmetry, since they involve the comparison of one or more candidate labelings against a single, unchanging ground truth.  Moreover, by contrast with the multitude of possible symmetric normalizations, the asymmetric measure we propose is the unique way to normalize the mutual information without introducing biases due to the normalization itself.

More broadly, the asymmetric measure is more informative than the conventional symmetric one.  Consider for instance the common situation where the groups or communities in one labeling are a subdivision of those in the other.  We might for example label individuals by the country they live in on the one hand and by the town or city on the other.  Then the more detailed labeling tells us everything there is to know about the coarser one, but the reverse is not true.  Telling you the city fixes the country, but not vice versa. Thus one could argue that the mutual information between the two eshould be asymmetric, and this type of asymmetry will be a key feature of the measures we study.

Both drawbacks of the standard mutual information---bias towards too many groups and dependence of the normalization on the candidate labeling---can be addressed simultaneously by using an asymmetric normalized reduced mutual information as defined in this paper.  In support of this approach we present an extensive comparison of the performance of this and other variants of the mutual information in network community detection tasks, generating a large number of random test networks with known community structure and a variety of structural parameters, and then attempting to recover the communities using popular community detection algorithms.  Within this framework, we find that conclusions about which algorithms perform best are significantly impacted by the choice of mutual information measure, and specifically that traditional measures erroneously favor algorithms that find too many communities, but our proposed measure does not. Code implementing our measure can be found at \url{https://github.com/maxjerdee/clustering-mi}.

\section{Results}
\label{sec:mutual-information}
Mutual information can be thought of in terms of the amount of information it takes to transmit a labeling from one person to another.  We represent a labeling or division of~$n$ objects into~$q$ groups as a vector of $n$ integer elements, each with value in the range~$1\ldots q$.  We assume there to be a ground-truth labeling, which we denote~$g$, with $q_g$ groups, and a candidate labeling for comparison~$c$ with $q_c$ groups, generated for instance by some sort of algorithm.  The mutual information~$I(c;g)$ between the two is the amount of information saved when transmitting the truth~$g$ if the receiver already knows the candidate~$c$. We can write this information as the total entropy of~$g$ minus the conditional entropy of $g$ given~$c$:
\begin{equation}
I(c;g) = H(g) - H(g|c).
\label{eq:I-Hg1-minus}
\end{equation}
Loosely, we say that $I(c;g)$ measures how much $c$ tells us about~$g$. Since the conditional information content $H(g|c)$ is nonnegative, the mutual information must satisfy inequalities
\begin{align}
I(c;g) \leq H(g), H(c),
\label{eq:MI-Inequalities}
\end{align}
which are saturated by
\begin{align}
I(g;g) = H(g), \qquad I(c;c) = H(c).
\end{align}
Carefully accounting for all of the constituent information costs, we can compute the mutual information from~\cite{JKN24}
\begin{align}
I(c;g) &= H(g) - H(g|c) \nonumber\\
    &= \log \frac{n!\prod_{rs} n_{rs}^{(gc)}!}{\prod_r n_r^{(c)}! \prod_s n_s^{(g)}!}  + \log \binom{n + q_g \alpha_g - 1}{q_g \alpha_g - 1}
    - \sum_{r=1}^{q_g} \log \binom{n_r^{(g)} + \alpha_g - 1}{\alpha_g - 1} \nonumber\\
    &\qquad - \sum_{s = 1}^{q_c} \log \binom{n_s^{(c)} + q_g \alpha_{g|c} - 1}{q_g \alpha_{g|c} - 1}
     + \sum_{r = 1}^{q_g}\sum_{s = 1}^{q_c} \log \binom{n_{rs}^{(gc)} + \alpha_{g|c} - 1}{\alpha_{g|c} + 1}.
\label{eq:I-Approx}
\end{align}
In this expression the quantity $n_r^{(g)}$ denotes the number of objects in labeling $g$ that belong to group~$r = 1\ldots q_g$, and similarly for~$n_s^{(c)}$.  The quantities~$n_{rs}^{(gc)}$ are the entries of the contingency table, and are equal to the number of objects that simultaneously belong to group~$s$ in the candidate labeling~$c$ and group~$r$ in the ground truth~$g$. Evaluating the sums and products over these table entries numerically, the value of $I(c;g)$ can be computed in time~$\Ord(q_c q_g)$. The parameter values~$\alpha_g$ and~$\alpha_{g|c}$ are chosen to optimize the transmission process using a fast line search, providing good compression for realistic labelings as discussed in~\cite{JKN24}.

The first term in Eq.~\eqref{eq:I-Approx} represents the information cost of transmitting the labelings given knowledge of the contingency table~$n_{rs}^{(gc)}$. The remaining terms account for the information costs associated with the transmission of the table itself.  In some cases the information required to transmit the $q_g \times q_c$ contingency table is significantly smaller than that needed for the $n$-vectors representing the labelings, and it is neglected in most treatments~\cite{NCY20, JKN24}, leading to the traditional mutual information measure, which we denote~$I_0$:
\begin{equation}
I_0(c;g) = \log \frac{n! \prod_{rs}n_{rs}^{(gc)}!}{\prod_r n_r^{(c)}! \prod_s n_s^{(g)}!}.
\label{eq:Mutual-Information}
\end{equation}
In other cases, however, the information cost of the contingency table can play a significant role and should be retained as in Eq.~\eqref{eq:I-Approx}, which yields the quantity known as the reduced mutual information.

As an example, suppose a community detection algorithm simply places each node in its own group, resulting in a candidate labeling $c = (1,2,3,\ldots)$.  No matter what the ground truth is, this choice of~$c$ clearly contains no information about it whatsoever, so we might expect the mutual information to be zero.  If we use the traditional mutual information, however, we get $I_0(c;g) = I_0(g;g)$, which is not merely nonzero, but implies that the candidate labeling is maximally informative, telling us everything about the ground truth.  This is as wrong as it could be---in fact the candidate tells us nothing at all.  The reduced mutual information, on the other hand, correctly says that $I(c;g) = 0$ (see Section~\ref{app:reduced-mutual-information} in the Supplementary Materials).

For the same reason, the traditional mutual information gives a maximum score to any candidate~$c$ which is a refinement of the true~$g$, i.e.,~every community in $c$ is a subset of a community in $g$. More generally, the traditional mutual information is biased towards labelings with too many groups~\cite{AP17, VEB10, NCY20,JKN24,LN16}.

A related shortcoming of the traditional mutual information is that for finite~$n$ even a random labeling~$c$ will have positive mutual information with respect to any ground truth in expectation: because the traditional mutual information is non-negative, fluctuations due to randomness will produce non-negative values only and hence their average will in general be positive~\cite{VEB10,Zhang15}.  This seems counterintuitive; we would expect the average value for a random labeling to be zero.

We can solve this problem by using another variant of the mutual information, which subtracts off the expected value, thereby making the average zero by definition.  To do this we must first specify how the expectation is defined---over what distribution of candidate labelings are we averaging?  The conventional choice is to take the uniform distribution over labelings that share the same group sizes~$n^{(c)}$ as the actual candidate~$c$.  This yields the adjusted mutual information of Vinh~\etal~\cite{VEB10}:
\begin{equation}
I_A(c;g) = I_0(c;g) - \bigl\langle I_0(c;g) \bigr\rangle_{\{c|n^{(c)}\}}\,,
\end{equation}
where the expectation $\langle\ldots\rangle$ is over the relevant ensemble.

The adjusted mutual information can also be derived in a fully information-theoretic manner, as described in~\cite{NCY20}.  There it is shown that the subtracted term $\langle I_0(c;g) \rangle_{\{c|n^{(c)}\}}$ is precisely equal to the average cost of transmitting the contingency table when labelings are drawn from the uniform distribution. However, this distribution heavily favors contingency tables with relatively uniform entries, simply because there are many more labelings that correspond to uniform tables than to non-uniform ones.  Real contingency tables, on the other hand, are often highly non-uniform, since applications of the mutual information focus on labelings that are somewhat similar to the ground truth (producing a non-uniform table).  In such cases, the average used in the adjusted mutual information puts most of its weight on configurations that are very different from those that occur in reality, making it a poor representation of true information costs.  The reduced mutual information considered in this paper, by contrast, deliberately allows for non-uniform tables by drawing them from a Dirichlet-multinomial distribution and we argue that this is a strong reason to favor it over the adjusted mutual information.  Nonetheless, in Section~\ref{sec:lfr} we give results using both reduced and adjusted mutual information, and find fairly similar outcomes in the two cases.

\subsection{Normalization of the mutual information}
\label{sec:normalization}
A fundamental difficulty with mutual information as a measure of similarity is that its range of values depends on the particular application, which makes it difficult to say when a value is large or small.  Is a mutual information of 10 a large value?  Sometimes it is and sometimes it isn't, depending on the context.  To get around this obstacle one commonly normalizes the mutual information so that it takes a maximum value of~1 when the candidate labeling agrees exactly with the ground truth.  There are a number of ways this can be achieved and, as we show here, they are not all equal.  In particular, some, including the most popularly used normalization, can result in biased results and should, in our opinion, be avoided.  In its place, we propose an alternative, unbiased normalized measure.

The most popular normalized measure, commonly referred to simply as the normalized mutual information, uses the plain mutual information~$I_0(c;g)$ as a base measure and normalizes it thus:
\begin{align}
\text{NMI}_0^{(S)}(c;g) &=  \frac{I_0(c;g)}{\frac{1}{2}[H_0(c) + H_0(g)]} 
 = \frac{I_0(c;g) + I_0(g;c)}{I_0(c;c) + I_0(g;g)}.
\label{eq:NMI-S-Definition}
\end{align}
This measure has a number of desirable features.  Because of the inequalities in~\eqref{eq:MI-Inequalities}, its value falls strictly between zero and one.  And since both the base measure and the normalization are symmetric under interchange of $c$ and~$g$, the normalized measure also retains this symmetry (hence the superscript $(S)$, for symmetric). 

\begin{figure}
\centering
\includegraphics[width=12cm]{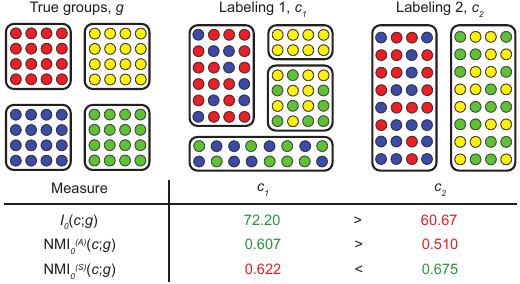}
\caption{Normalization can impact which labeling is preferred by a mutual information measure. In this example 64 objects are split into four equally sized true groups~$g$, denoted by their color. Against this ground truth, we compare two candidate labelings, $c_1$ and~$c_2$. The standard unnormalized mutual information~$I_0(c;g)$ of Eq.~\eqref{eq:Mutual-Information} reports that labeling~$c_1$ shares more bits of information with the ground truth than does~$c_2$. By definition, the asymmetrically normalized mutual information~$\text{NMI}^{(A)}$ will always agree with the unnormalized measure, as in this example. The symmetrically normalized~$\text{NMI}^{(S)}$, on the other hand, is biased in favor of simpler labelings, which makes it prefer labeling~$c_2$ in this case.}
\label{fig:nmivmifigure}
\end{figure}

Equation~\eqref{eq:NMI-S-Definition} is not the only normalization that achieves these goals.  Equation~\eqref{eq:MI-Inequalities} implies that
\begin{align}
I_0(c;g) &\leq \min(I_0(c;c),I_0(g;g)) \nonumber\\
    &\leq \sqrt{I_0(c;c) I_0(g;g)}     \nonumber\\
    &\leq \max(I_0(c;c),I_0(g;g)),
\label{eq:Symmetric-Normalization-Denoms}
\end{align}
which gives us three more options for a symmetric denominator in the normalized measure.  The arithmetic mean in Eq.~\eqref{eq:NMI-S-Definition}, however, sees the most use by far~\cite{LF09,OLC11,YAT16,fortunato2016community}.

We can extend the notion of symmetric normalization to any other base measure of mutual information~$I_X(c;g)$ satisfying the inequality Eq.~\eqref{eq:MI-Inequalities}, such as adjusted or reduced mutual information~\cite{JKN24}, by writing
\begin{equation}
\text{NMI}_X^{(S)}(c;g) = \frac{I_X(c;g)+I_X(g;c)}{I_X(c;c)+I_X(g;g)}. \label{eq:Generic-M-S-Definition}
\end{equation}
All such measures, however, including the standard measure of Eq.~\eqref{eq:NMI-S-Definition}, share a crucial shortcoming, that the normalization depends on the candidate labeling~$c$ and hence that the normalized measure can prefer a different candidate labeling to the base measure purely because of the normalization.

Figure~\ref{fig:nmivmifigure} shows an example of how this can occur. In this example 64 objects are split into four equally sized groups in the ground-truth labeling~$g$, against which we compare two candidate labelings, $c_1$~and~$c_2$. Neither candidate is perfect.  Labeling~$c_1$ identifies the correct number of groups but misplaces certain objects, while labeling~$c_2$ contains only two groups, each an aggregation of a pair of true groups. Under the \hbox{unnormalized} mutual information of Eq.~\eqref{eq:Mutual-Information}, $c_1$~receives a higher score than~$c_2$, but under the normalized measure of Eq.~\eqref{eq:NMI-S-Definition} the reverse is true. This behavior is due to the difference in entropy~$I_0(c;c)$ between the two candidate divisions. Since the candidate entropy appears in the denominator of the symmetric normalization, Eq.~\eqref{eq:NMI-S-Definition}, the symmetric measure favors simple labelings with smaller entropy. In this example, candidate~$c_2$ has only two groups, which results in a smaller entropy and a bias in its favor.

We argue that the unnormalized measure is more correct on this question, having a direct justification in terms of information theory.  The purpose of the normalization is merely to map the values of the measure onto a convenient numerical interval, and should not change the outcome as it does here.  Moreover, different symmetric normalizations can produce different results.  For instance, if one normalizes by $\max(I_0(c;c),I_0(g;g))$ in Fig.~\ref{fig:nmivmifigure} then candidate~$c_1$ is favored in all cases.

\begin{figure}
\centering
\includegraphics[width=11cm]{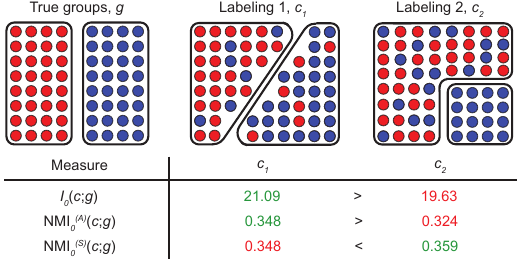}
\caption{Another example where the choice of normalization affects which labeling is preferred, this one with the same number of groups in both divisions. Although both candidates~$c_1$ and~$c_2$ correctly split the objects into two groups, candidate labeling~$c_1$ scores higher with respect to both the unnormalized and asymmetrically normalized mutual informations while the symmetric normalization prefers~$c_2$.}
\label{fig:nmivmifigure2}
\end{figure}

The choice of normalization can also change conclusions even when the number of groups does not vary.  Consider the example shown in Fig.~\ref{fig:nmivmifigure2}, in which the ground truth~$g$ splits the objects into two equally sized groups. Candidate~$c_1$ also has two groups of equal size, although each is slightly polluted with the objects of the opposite group. Candidate~$c_2$ has two groups of different sizes, a small group with one color only and a larger group with objects of both colors. Although both labelings have the same number of groups, candidate~$c_2$ has a smaller entropy~$I_0(c;c)$, since it is easier to communicate group identity within~$c_2$---most objects belong to the larger group. As a result, the symmetrically normalized mutual information prefers~$c_2$, while the unnormalized measure prefers~$c_1$. 

These issues are unavoidable when using a symmetric normalization scheme.  In any such scheme the normalization must depend on both~$c$ and~$g$ and hence can vary with the candidate labeling.  However, if we drop the requirement of symmetry then we can normalize in a way that avoids these issues.  We define the asymmetric normalization of any base measure $I_X$~as
\begin{equation}
\text{NMI}_X^{(A)}(c;g) = \frac{I_X(c;g)}{I_X(g;g)}. \label{eq:General-M-A-Definition}
\end{equation}
This definition still gives~$\text{NMI}_X^{(A)}(g;g) = 1$, but now the normalization factor in the denominator has no effect on choices between candidate labelings, since it is independent of~$c$.  In fact, Eq.~\eqref{eq:General-M-A-Definition} is the only way to normalize such that $I_X^{(A)}(g;g) = 1$ while simultaneously ensuring that the preferred candidate is always the same as for the base measure.  Thus this measure also removes any ambiguity about how one should perform the normalization. Loosely, this asymmetrically normalized mutual information measures how much~$c$ tells us about~$g$ as a fraction of all there is to know about~$g$. Asymmetrically normalized measures have appeared in other contexts under the name ``coefficient of constraint'' or ``coefficient of uncertainty''~\cite{CDT70, Theil92, Effenberger13}, although symmetric normalizations are almost universally used in the model validation context, where they introduce the bias we discuss here.

The amount of bias inherent in the symmetrically normalized measure when compared with the asymmetric one can be quantified by the ratio between the two:
\begin{align}
{\text{NMI}_X^{(S)}(c;g)\over\text{NMI}_X^{(A)}(c;g)}
  &= {[I_X(c;g)+I_X(g;c)]/[I_X(c;c)+I_X(g;g)]\over
  I_X(c;g)/I_X(g;g)} \nonumber\\
  &= \frac{1 + I_X(g;c)/I_X(c;g)}{1 + I_X(c;c)/I_X(g;g)}.
\end{align}
If the base measure~$I_X$ is itself symmetric (which it is, either exactly or approximately, for all the measures we consider), then this simplifies further to
\begin{equation}
{\text{NMI}_X^{(S)}(c;g)\over\text{NMI}_X^{(A)}(c;g)}
  = {H(g)\over\frac12[H(g)+H(c)]}.
  \label{eq:symmetric-normalization-bias}
\end{equation}
Values of this quantity below (above)~1 indicate that the symmetric measure is biased low (high).  Thus, for instance, complex candidate labelings~$c$ that have higher entropy than the ground truth will result in a symmetric measure whose values are too low.  As discussed in the introduction, traditional mutual information measures are particularly problematic when the candidate is a refinement of the ground truth, meaning the candidate groups are subsets of the ground-truth groups.  In that case the traditional measure returns values that are too high. Equation~\eqref{eq:symmetric-normalization-bias} implies that, to some extent, the symmetric normalization will correct this issue: a candidate~$c$ that takes the form of a refinement of~$g$ will have $H(c)>H(g)$, which will lower the value of the symmetrically normalized mutual information.  This may in part explain why these biases have been overlooked in the past: two wrongs have conveniently canceled to (approximately) make a right.

We argue, however, that this is not the best way to address the problem and that the correct approach is instead to use the reduced mutual information.  The reduced mutual information also corrects for the case where one labeling is a refinement of the other, but does so in a more principled manner that directly addresses the root cause of the problem, rather than merely penalizing complex candidate labelings in an ad hoc manner as a side-effect of normalization.

We will see examples of these effects in Section~\ref{sec:results}, where we apply the various measures to community detection in networks and find that indeed the traditional symmetrically normalized mutual information is biased.  In some cases it is fortuitously biased in the right direction, although it is still problematic in some others.

\begin{figure}
\centering
\includegraphics[width=14cm]{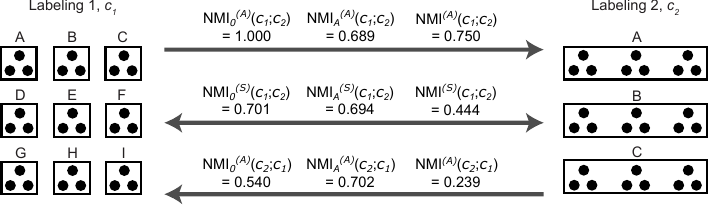}
\caption{Values of the various normalized mutual information measures between two example partitions of the same set of 27 objects.  The left partition~$c_1$ divides the objects into nine groups of three objects each and is a refinement of the right partition~$c_2$, which divides them into only three groups.  The arrows indicate the direction of the comparison and the accompanying notations give the values of the mutual information measures.  All three asymmetrically normalized measures capture the intuition that the partition $c_1$ tells us more about $c_2$ than $c_2$ tells us about $c_1$. }
\label{fig:asymmetric-example}
\end{figure}

An obvious downside of asymmetric normalization is the loss of the symmetry in the final measure.  In the most common applications of normalized mutual information, where labelings are evaluated against a ground truth, an inherently asymmetric situation, the asymmetric measure makes sense, but in other cases the lack of symmetry can be undesirable.  Embedding and visualization methods that employ mutual information as a similarity measure, for example, normally demand symmetry~\cite{PLC17}.  And in cases where one is comparing two candidate labelings directly to one another, rather than to a separate ground truth, a symmetric measure may be preferable.  Even in this latter case, however, the asymmetric measure may sometimes be the better choice, as discussed in the introduction.  For instance, when one labeling~$c_1$ is a refinement of the other~$c_2$, the information content is inherently asymmetric: $c_1$~says more about $c_2$ than $c_2$ does about~$c_1$.  An explicit example of this type of asymmetry is shown in Figure~\ref{fig:asymmetric-example}, where we consider two labelings of~27 objects. The left labeling~$c_1$ is a detailed partition of the objects into nine small groups while the right labeling~$c_2$ is a coarser partition into only three groups, each of which is an amalgamation of three of the smaller groups in~$c_1$.  Because of this nested relationship, it is relatively easy to transmit~$c_2$ given knowledge of~$c_1$ but more difficult to do the reverse.  This imbalance is reflected in the asymmetric normalized mutual information values in each direction (top and bottom arrows in the figure), but absent from the symmetric version (middle row).

Combining the benefits of asymmetric normalization and the reduced mutual information, we advocate in favor of the asymmetrically-normalized reduced mutual information defined by
\begin{equation}
\text{NMI}^{(A)}(c;g) = \frac{I(c;g)}{I(g;g)},
\end{equation}
where the mutual information $I(c;g)$ is quantified as in Eq.~\eqref{eq:I-Approx}.  This measure correctly accounts for the information contained in the contingency table, returns a negative value when $c$ is unhelpful for recovering the ground truth, returns~1 if and only if $c=g$, and always favors the same labeling as the unnormalized measure.  The traditional normalized mutual information possesses none of these desirable qualities.

\subsection{Copurchasing network of books}
\label{sec:results}
As an example of the performance of the various measures discussed above, in this section we use them to score the output of algorithms for network community detection. We consider a selection of networks with known true groups and attempt to recover those groups using various standard community detection algorithms, quantifying the accuracy of recovery with six different measures: the symmetrically and asymmetrically normalized versions of the traditional mutual information, the adjusted mutual information, and the reduced mutual information. 

\begin{figure}[t]
\centering
\includegraphics[width=11.5cm]{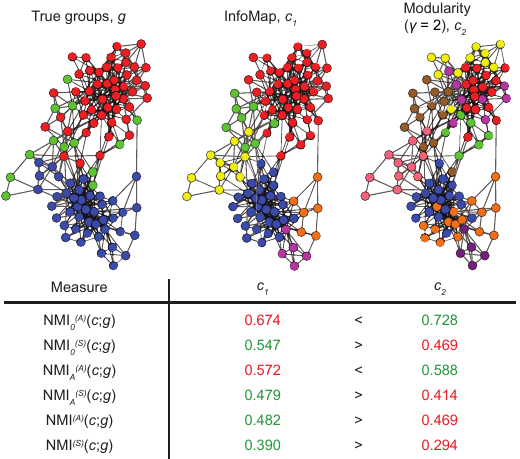}
\caption{Performance of the InfoMap and $\gamma = 2$ modularity maximization algorithms in recovering the true groups in the copurchasing network of political books~\cite{polbooks}. True and recovered groups are indicated by node color, and similarity between the two is assessed using the mutual information in its traditional~$I_0$, adjusted~$I_A$, and reduced~$I$ variants, normalized asymmetrically~$\text{NMI}^{(A)}$ and symmetrically~$\text{NMI}^{(S)}$. The choices of both the base measure and the normalization impact which algorithm is preferred.}
\label{fig:polbooks-example}
\end{figure}

As a first example, Fig.~\ref{fig:polbooks-example} shows a network compiled by V.~Krebs of 105 books about US politics published around the time of the 2004 US presidential election.  The true groups~$g$ in this case represent the political lean of each book's content: liberal, conservative, or neutral~\cite{polbooks}, as determined by Krebs.  Network edges connect pairs of books frequently purchased together on Amazon.com, and such books typically are of the same political bent, either both liberal or both conservative, so that the political positions form communities of connected network nodes.  Community detection algorithms, which aim to find such communities, should therefore be able to recover political positions, at least approximately, from the network of connections alone. The figure shows the results of applying two of the most popular such algorithms, InfoMap and modularity maximization (with resolution parameter $\gamma=2$), but we find that these algorithms identify quite different partitions of the network, with InfoMap finding six groups where modularity finds nine. We use the normalized mutual information in its various forms to assess how closely each division matches the ground truth.

We find that both the base mutual information measure~$I_X$ and its normalization considerably impact our assessments.  For instance, the asymmetrically normalized traditional mutual information gives a score of $\text{NMI}_0^{(A)}(c_2;g) = 0.728$, out of a maximum of~1, for the partition found by modularity maximization, which is over twice the score of $\text{NMI}^{(S)}(c_2;g) = 0.294$ given to the same partition by the symmetrically normalized reduced measure. 

The choice of mutual information measure can also change which algorithm is preferred. For the traditional unreduced base measure~$I_0$, the asymmetrically normalized mutual information favors the groups found by modularity while the symmetric version favors the simpler partition found by InfoMap.  The same pattern holds if the adjusted~$I_A$ is used. The reduced measure~$I$, however, prefers InfoMap regardless of normalization, although by a narrower margin in the asymmetric case. As discussed in the previous section, we favor the asymmetrically normalized reduced measure on formal grounds, and view the (often large) deviations from its assessment as unnecessary biases.

\subsection{LFR benchmark networks}
\label{sec:lfr}
In one sense the previous example is unusual: there are relatively few cases like this of networks with known ground-truth communities to compare against. To get a more comprehensive picture of algorithm performance, therefore, researchers have turned to synthetic benchmark networks like those generated by the popular Lancichinetti-Fortunato-Radicchi (LFR) graph model~\cite{LFR08}, which creates networks with known community structure and realistic distributions of node degrees and group sizes. A~number of studies have been performed in the past to test the efficacy of community detection algorithms on LFR benchmark networks~\cite{LF09,OLC11,YAT16,fortunato2016community}, but using only the symmetrically normalized, non-reduced mutual information as a similarity measure.  Our results indicate that this measure can produce biased outcomes and we recommend the asymmetric reduced mutual information instead. 

The LFR model contains a number of free parameters that control the size of the networks generated, their degree distribution, the distribution of community sizes, and the relative probability of within- and between-group edges.  (See Section~\ref{app:benchmarking} in the Supplementary Materials for details of the LFR generative process.)  We find that the distributions of degrees and community sizes do not significantly impact the relative performance of the various algorithms tested and that performance differences are driven primarily by the size~$n$ of the networks and the mixing  parameter~$\mu$ that controls the ratio of connections within and between groups, so our tests focus on performance as a function of these parameters.

\begin{figure*}
\centering
\includegraphics[width=\textwidth]{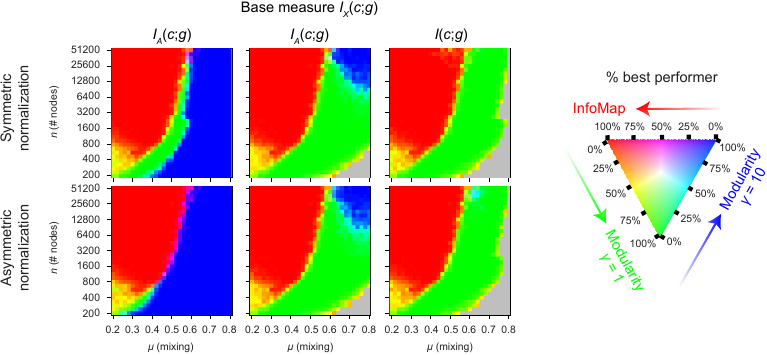}
\caption{Performance of three community detection algorithms: InfoMap (red) and modularity maximization with $\gamma=1$ (green) and $\gamma=10$ (blue)---see key on right.  The colors in each panel of this figure indicate which of the three is best able to find the known communities in a large set of LFR benchmark networks, according to the six mutual information measures we consider. Mixtures of red, green, and blue denote the proportions of test cases in which each algorithm performs best.  Regions in gray indicate parameter values for which no algorithm achieved a positive mutual information score.}
\label{fig:RGB-Comparison}
\end{figure*}

\subsection{Comparison between variants of the  mutual information}
\label{mutual-information-comparison}
Figure~\ref{fig:RGB-Comparison} summarizes the relative performance of the various mutual information measures in our tests.  In this set of tests we limit ourselves, for the sake of clarity, to the top three community detection algorithms---InfoMap and the two variants of modularity maximization---and measure which of the three returns the best results according to each of our six mutual information measures, as a function of network size~$n$ and the mixing parameter~$\mu$.  Each point in each of the six panels is color-coded with some mix of red, green, and blue to indicate in what fraction of cases each of the algorithms performs best according to each of the six measures and, as we can see, the results vary significantly among measures.  An experimenter trying to choose the best algorithm would come to substantially different conclusions depending on which measure they use.

One consistent feature of all six mutual information measures is the large red area in each panel of Fig.~\ref{fig:RGB-Comparison}, which represents the region in which the InfoMap algorithm performs best.  Regardless of the measure used, InfoMap is the best performer on networks with low mixing parameter (i.e.,~strong community structure) and relatively large network size.  For higher mixing (weaker structure) or smaller network sizes, modularity maximization does better.  Which version of modularity is best, however, depends strongly on the mutual information measure.  The traditional symmetrically normalized mutual information (top left panel) mostly favors the version with a high resolution parameter of $\gamma=10$ (blue), but the asymmetric reduced measure for which we advocate (bottom right) favors the version with $\gamma=1$ (green).  (The regions colored gray in the figure are those in which no algorithm receives a positive mutual information score and hence all algorithms can be interpreted as failing.)

These results raise significant doubts about the traditional measure. Consider the lower right corner of each plot in Fig.~\ref{fig:RGB-Comparison}, which is the regime of small network sizes~$n$ and high mixing~$\mu$ so that the community signal is weak and the noise is high. Here, the adjusted and reduced mutual informations indicate that all algorithms are failing. This is expected: for very weak community structure all detection algorithms are expected to show a ``detectability threshold'' beyond which they are unable to identify any communities~\cite{DKMZ11a,Massoulie14,MNS15}.  The standard normalized mutual information, on the other hand, claims to find community structure in this regime using the $\gamma=10$ version of modularity maximization.  This occurs because the $\gamma=10$ algorithm finds many small communities and, as discussed in Section~\ref{sec:mutual-information}, a labeling with many communities, even completely random ones, is accorded a high score by a non-reduced mutual information. The presence of this effect is demonstrated explicitly in the Supplementary Materials, Section~\ref{app:number-of-groups} for the case of modularity maximization with~$\gamma = 10$, and this offers a clear reason to avoid the standard measure. 

\begin{figure*}
\centering
\includegraphics[width=14cm]{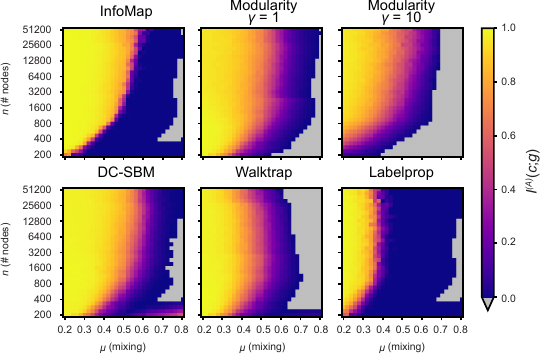}
\caption{Performance of each of the six community detection algorithms considered here in identifying the known communities in a large set of LFR benchmark networks, as quantified by the asymmetrically normalized reduced information measure $I^{(A)}(c;g)$ proposed in this paper. Gray areas indicate parameter values for which~$\text{NMI} < 0$, so that the conditional encoding is less efficient than the direct encoding.}
\label{fig:RMI-A-Performances}
\end{figure*}

The bottom left panel in Fig.~\ref{fig:RGB-Comparison} shows results for the asymmetrically normalized version of the traditional mutual information, which gives even worse results than the symmetric version, with hardly any region in which the $\gamma=1$ version of modularity maximization outperforms the $\gamma=10$ version.  This behavior arises for the reasons discussed in Section~\ref{sec:normalization}---the bias inherent in the symmetric normalization fortuitously acts to partially correct the errors introduced by neglecting the information content of the contingency table.  The asymmetric normalization eliminates this correction and hence performs more poorly.  The correct solution to this problem, however, is not to use a symmetric normalization, which can bias outcomes in other ways as we have seen, but rather to adopt a reduced mutual information measure.

Finally, comparing the middle and right-hand columns of Fig.~\ref{fig:RGB-Comparison}, we see that the results for the adjusted and reduced mutual information measures are quite similar in these tests, although there are some differences.  In particular, the adjusted measure appears to find more significant structure for higher mixing than the reduced measure.  This occurs because, as discussed in Section~\ref{sec:mutual-information} and Ref.~\cite{NCY20}, the adjusted measure encodes the contingency table in a way that is optimized for more uniform tables than the reduced measure, and thus penalizes uniform tables less severely, leading to overestimates of the mutual information in the regime where detection fails---in this regime the candidate and ground-truth labelings are uncorrelated which results in a uniform contingency table.  This provides further evidence in favor of using a reduced mutual information measure.

\subsection{Comparison between community detection algorithms}
\label{sec:algorithms}
Settling on the asymmetrically normalized reduced mutual information as our preferred measure of similarity, we now ask which community detection algorithm or algorithms perform best according to this measure?  We have already given away the answer---InfoMap and modularity maximization get the nod---but here we give evidence for that conclusion.

Figure~\ref{fig:RMI-A-Performances} shows results for all six algorithms listed in the Supplementary Materials, Section~\ref{app:algorithms}.  Examining the figure, we see that in general the best-performing methods are InfoMap, traditional modularity maximization with $\gamma=1$, and the inference method using the degree-corrected stochastic block model.  Among the algorithms considered, InfoMap achieves the highest mutual information scores for lower values of the mixing parameter~$\mu$ in the LFR model, but fails abruptly as $\mu$ increases, so that beyond a fairly sharp cutoff around $\mu=0.5$ other algorithms do better.  As noted by previous authors~\cite{ACM22}, Info\-Map's specific failure mode is that it places all nodes in a single community and this behavior can be used as a simple indicator of the failure regime.  In this regime one must use another algorithm.  Either the modularity or inference method are reasonable options, but modularity has a slight edge, except in a thin band of intermediate $\mu$ values which, in the interests of simplicity, we choose to ignore.  (We discuss some caveats regarding the relationship between the degree-corrected stochastic block model and the LFR benchmark in Section~\ref{app:benchmarking} of the Supplementary Materials.)

Thus---always assuming the LFR benchmark is a good test of performance---our recommendations for the best community detection algorithm are relatively straightforward.  If we are in a regime where InfoMap succeeds, meaning it finds more than one community, then one should use InfoMap.  If not, one should use standard modularity maximization with $\gamma=1$.  That still leaves open the question of how the modularity should be maximized.  In our studies we find the best results with simulated annealing, but simulated annealing is computationally expensive.  In regimes where it is not feasible, we recommend using the Leiden algorithm instead.  (Tests using other computationally efficient maximization schemes, such as the Louvain and spectral algorithms, generally performed less well than the Leiden algorithm.)

\section{Discussion}
In this paper we have examined the performance of a range of mutual information measures for comparing labelings of objects in classification, clustering, or community detection applications.  We argue that the commonly used normalized mutual information is biased in two ways: (1)~because it ignores the information content of the contingency table, which can be large, and (2)~because the symmetric normalization it employs introduces spurious dependence on the labeling.  We argue in favor of a different measure, an asymmetrically normalized version of the reduced mutual information, which rectifies both of these shortcomings.

To demonstrate the effects of using different mutual information measures, we have presented results of an extensive set of numerical tests on popular network community detection algorithms, as evaluated by the various measures we consider.  We find that conclusions about which algorithms are best depend substantially on which measure we use.

\section{Data availability}
All datasets analyzed here have been deposited in the Zenodo database under accession code \href{https://doi.org/10.5281/zenodo.17211478}{10.5281/zenodo.17211478}, except for the synthetic LFR benchmark networks, which were randomly generated using the open-source NetworkX package as described in the Supplementary Materials, Section~\ref{app:benchmarking}.

\subsection*{Code availability}
Python code implementing the mutual information measures discussed in this paper is available at  \url{https://github.com/maxjerdee/clustering-mi} or~\cite{JKN25Zenodo}. 

\subsection*{Acknowledgments}
The authors thank Samin Aref for useful comments and feedback.  This work was supported in part by the US National Science Foundation under grants DMS--2005899 and DMS--2404617 (MN), the National Science Foundation of China through Young Scientist Fund Project grant 12405044 (AK), and computational resources provided by the Advanced Research Computing initiative at the University of Michigan.

\subsection*{Author contributions}
\review{All authors (MJ, AK, and MN) conceptualized the research and wrote and edited the paper.  Code development and data analysis were performed by MJ.  All authors have read the final manuscript.}

\subsection*{Competing interests}
\review{The authors declare no competing interests.}

\appendix
\section{Benchmarking}\label{app:benchmarking}
In this appendix we give some additional details of our community detection tests.

\subsection{Numbers of groups}
\label{app:number-of-groups}

\begin{figure}
\centering
\includegraphics[width=8cm]{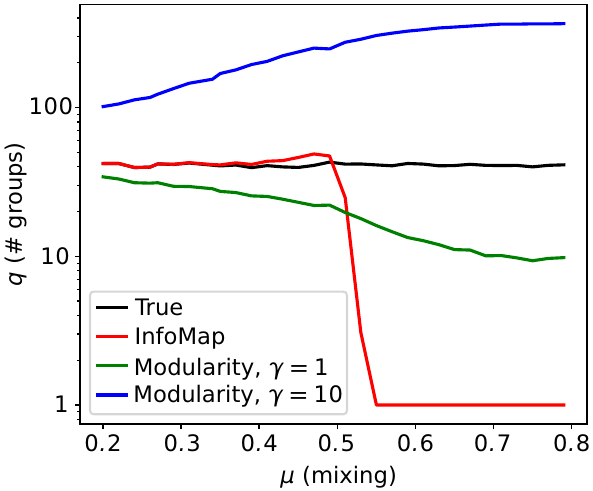}
\caption{The number of groups inferred by each of the six algorithms in Fig.~\ref{fig:RGB-Comparison} for LFR benchmark networks with $n = 3200$ nodes and a range of values of the mixing parameter~$\mu$.  The ground-truth number of groups is shown in black.  The InfoMap algorithm (red) generates an accurate number of groups for values of $\mu$ up to about 0.5, but beyond this point it erroneously places all nodes in a single group.  Standard modularity maximization with resolution parameter $\gamma=1$ (green) underestimates the number of groups, perhaps because of the resolution limit on the detection of small groups, but not as severely as the number is overestimated when $\gamma = 10$ (blue).}
\label{fig:comm-q-plot}
\end{figure}

As discussed in Section~\ref{mutual-information-comparison}, the traditional (non-reduced) mutual information favors the~$\gamma = 10$ generalized modularity maximization algorithm over other algorithms across many of the tests reported in Fig.~\ref{fig:RGB-Comparison}. However, a closer inspection of the partitions found by this algorithm reveals that it drastically overestimates the number of communities. The number of groups inferred by each algorithm for each value of the mixing parameter $\mu$ is shown for $n = 3200$ in Fig.~\ref{fig:comm-q-plot}.  The traditional mutual information has a bias towards labelings with an excessive number of groups, which causes it to prefer the~$\gamma = 10$ algorithm in this regime, while the reduced and adjusted mutual information measures prefer simple modularity maximization ($\gamma=1$).

\subsection{Community detection algorithms}
\label{app:algorithms}
We assess the performance of six common community detection algorithms on synthetic network examples.  We use the algorithm implementations from the \verb|igraph| library~\cite{CN06}, except for the inference method, for which we use the \verb|graph-tool| library~\cite{P14}.  The six algorithms are as follows.
\begin{enumerate}
\item \textbf{InfoMap:} InfoMap is an information theoretic approach that defines a compression algorithm for encoding a random walk on a network, based on which communities the walk passes through~\cite{RB08}.  Different community labelings give rise to more or less efficient compression, as quantified by the so-called map equation, and the labeling with the highest efficiency is considered the best community division. 
\item \textbf{Modularity maximization:} Modularity is a quality function for network community divisions equal to the fraction of edges within communities minus the expected such fraction if edge positions are randomized while preserving node degrees.  The labeling with the highest modularity is considered the best community division.  Exact modularity maximization is NP-hard and usually intractable, but modularity can be maximized approximately using various heuristics, of which the most popular are agglomerative methods such as the Louvain and Leiden algorithms~\cite{BGLL08, TWV19}, spectral methods~\cite{Newman06b}, and simulated annealing~\cite{GSA04,MAD05,RB06a}.  In our tests we use simulated annealing where computationally feasible and the Leiden algorithm otherwise, these approaches giving the most consistent maximization of the modularity.
\item \textbf{Modularity with a resolution parameter:} Standard modularity maximization is known to suffer from a ``resolution limit''---it cannot detect communities smaller than a certain threshold size~\cite{FB07}.  This can be remedied using a variant of modularity that includes a resolution parameter~$\gamma$ such that higher values of $\gamma$ cause the algorithm to prefer smaller communities~\cite{RB06a}.  Standard modularity maximization corresponds to $\gamma=1$, but for comparison we also conduct tests with $\gamma=10$ using the Leiden algorithm.
\item \textbf{Statistical inference:} Another popular approach to community detection makes use of model fitting and statistical inference.  In this context the most commonly fitted model is the degree-corrected stochastic block model~\cite{KN11a}, which can be fitted using Bayesian methods to find the best community division~\cite{Peixoto17}.
\item \textbf{Walktrap:} Walktrap is an agglomerative algorithm in which initially separate nodes are iteratively combined into progressively larger communities in order from strongest to weakest connections, where strength is quantified in terms of the time for a random walk to reach one node from another~\cite{PL05}.
\item \textbf{Labelprop:} The label propagation or ``labelprop'' algorithm initially places every node in its own community then iteratively updates the labels of randomly chosen nodes by majority vote among their network neighbors, breaking ties at random~\cite{RRS07}.
\end{enumerate}
As discussed in Section~\ref{sec:algorithms}, all of these algorithms perform reasonably well, but the best performers in our tests are InfoMap and the two variants of modularity maximization.

\subsection{Results for the traditional symmetric normalized mutual information} 
\label{app:alternative-results}
\begin{figure*}
\centering
\includegraphics[width=14cm]{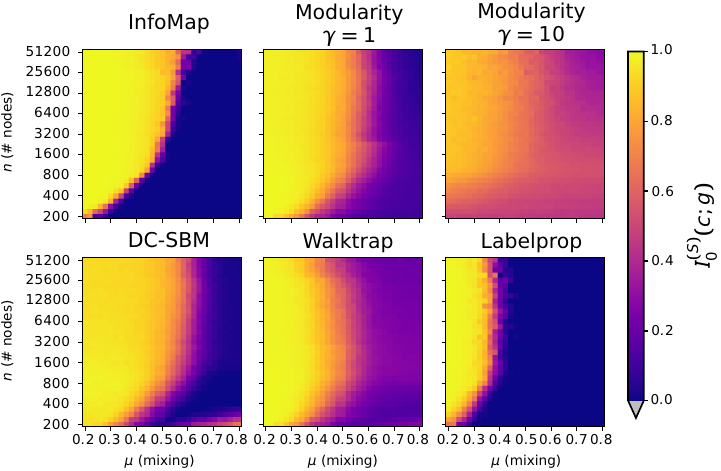}
\caption{Performance of each of the six community detection algorithms considered here, as quantified by the conventional symmetrically normalized non-reduced mutual information~$I_0^{(S)}(c;g)$.}
\label{fig:NMI-S-Performances}
\end{figure*} 
Figure~\ref{fig:NMI-S-Performances} shows the performance of the same six community detection methods as in Fig.~\ref{fig:RMI-A-Performances}, but measured using the standard, symmetrically normalized, non-reduced mutual information, which we denote by~$I_0^{(S)}(c;g)$.  By this measure, many of the methods appear to perform implausibly well, far beyond the detectability threshold visible in Fig.~\ref{fig:RMI-A-Performances}, in the regime where all methods should by rights fail.  Note in particular the high scores achieved by the generalized modularity with $\gamma = 10$ by virtue of the excessive number of groups it generates.

\subsection{LFR network generation}
\label{app:benchmarking-details}
The LFR networks we use for benchmarking are generated using the procedure described in~\cite{LFR08}, which we summarize here.
\begin{enumerate}
    \item \textbf{Fix the number of nodes~$n$ and mixing parameter~$\mu$}.  In our studies we use node counts in the range~$n \in [200,51200]$.  The parameter $\mu$ controls the relative number of edges within and between communities.  For small $\mu$ there are many more edges within communities than between them, which makes the communities easy to detect.  But as $\mu$ gets larger there are more edges between communities and detection becomes more difficult in a manner reminiscent of the detectability threshold in the standard stochastic block model~\cite{DKMZ11a,DKMZ11b}, although there is no known analytic detectability threshold for the LFR model.  In our studies we use values of $\mu$ in the range $[0.2,0.8]$.
    \item \textbf{Draw a degree sequence} from a power-law distribution with exponent~$\tau_1$.  Many networks have power-law degree distributions, typically with exponents between 2 and~3~\cite{Caldarelli07}, and the LFR model exclusively uses power-law distributions.  We use~$\tau_1 = 2.5$, with average degree~$\langle k \rangle = 20$ and maximum degree (which depends on graph size) $k_{\text{max}} = n/10$.  Empirically, however, our results do not seem to be very sensitive to these choices.
    \item \textbf{Draw a set of community sizes} from a power-law distribution with exponent~$\tau_2$.  Many networks are also found to have community sizes that approximately follow a power law, with typical exponents in the range from 1 to~2~\cite{Guimera03,CNM04,PDFV05,LFR08}.  We use $\tau_2 = 1.5$ with a minimum community size of~$s_{\text{min}} = 20$ in all cases, while the maximum community size is set to~$s_{\text{max}} = \max\left(n/10,100\right)$. Again, results were not particularly sensitive to these choices, provided they produce a valid distribution at all. 
    \item \textbf{Assign each node to a community} randomly, one node at a time, ensuring that the community chosen is large enough to support the added node's intra-community degree, given by $(1 - \mu) k$ where $k$ is the total degree. 
    \item \textbf{Rewire the edges} attached to each node while preserving the node degrees, until the fraction of edges running between nodes in different communities is approximately $\mu$.
\end{enumerate}
The parameter values above are similar to those used for instance in~\cite{YAT16}.  As in that study, we find that algorithm performance is dictated primarily by the parameters $n$ and~$\mu$, so it is these parameters that are varied our summary figures.

The LFR model is similar to a special case of the degree-corrected stochastic block model (DC-SBM)~\cite{KN11a}, and hence one might expect that inference-based community detection methods employing the latter model would perform well, perhaps even optimally, on LFR networks.  Specifically, in the limit of an infinite number of sampled networks and perfect optimization of each community detection method, the final performance measure for any algorithm is given by the expectation value of the similarity $M(g,h[A])$ between the ground truth LFR partition~$g$ and the partition $h[A]$ of the LFR network~$A$ inferred using the algorithm, where the expectation is taken over the ensemble $P(A,g|\theta)$ of LFR networks and partitions $A,g$ generated using parameters $\theta$ (meaning $\mu$, $\tau_1$, $\tau_2$, etc.).  By using the LFR benchmark with parameters~$\theta$ to compare the performance of community detection algorithms, we are therefore implicitly defining the ``best'' algorithm to be the one whose corresponding function $h[A]$ optimizes $\sum_{A,g} P(A,g|\theta) M(g,h[A])$.  If the similarity measure we choose is the ``all or nothing'' error function $M(g,c) = \delta(g,c)$, then the optimal community detection algorithm is trivially the one with
\begin{equation}
    h[A] = \argmax_g P(A,g|\theta) = \argmax_g P(g|A,\theta).
\end{equation}
In other words, the optimal algorithm simply performs maximum a posteriori estimation under the model from which the network was generated.  There is no explicit formula for the posterior probability under the LFR model, but to the extent that it is a special case of the DC-SBM, we might expect the DC-SBM (with appropriate priors) to give optimal results~\cite{peixoto2021revealing}.

The LFR model, however, is not precisely a special case of the DC-SBM.  In particular, the DC-SBM normally assumes a uniform distribution over community sizes, where the LFR model assumes a power law.  Moreover, we are not using the crude all-or-nothing error function: our entire purpose in this paper is to develop mutual information measures that aggregate and weigh different modes of error in a sensible fashion.  These differences, it appears, are enough to ensure that the DC-SBM does not perform the best in our testing. 

Regardless, we emphasize that our use of the LFR benchmark in our analysis is simply for consistency with previous studies of network community detection methods~\cite{LFR08,LF09,YAT16}.  The justification for our proposed similarity measure, on the other hand, is chiefly its theoretical merit over the conventional (symmetric, non-reduced) normalized mutual information, and is independent of the use of the LFR (or any other) benchmark.

\section{Bias towards large numbers of groups in the traditional mutual information}
\label{app:reduced-mutual-information}

To shed light on why the traditional mutual information of Eq.~\eqref{eq:Mutual-Information} is biased towards an excessive number of groups and how the reduced mutual information of Eq.~\eqref{eq:I-Approx} corrects this, consider the (extreme) example in which every object is placed in a group on its own, producing a candidate labeling~$c = (1,\ldots,n)$.  Whatever the ground truth labeling $g$ is, this candidate labeling clearly provides no information about it whatsoever, so we expect the mutual information to be zero.  But this is not what we find.  The contingency table in this case~is
\begin{equation}
n_{rs}^{(gc)} = \biggl\lbrace\begin{array}{ll}
1 & \quad\mbox{if $r = g_s$,}   \\
0 & \quad\mbox{otherwise,} \\
\end{array}
\label{eq:n12Example}
\end{equation}
so the conventional mutual information of Eq.~\eqref{eq:Mutual-Information} is
\begin{equation}
I_0(c;g) = \log \frac{n!}{\prod_s n_s^{(g)}!} = H_0(g). 
\end{equation}

This answer is not merely wrong; it is maximally so.  The mutual information should take its minimum value of zero, but instead it takes the value~$H_0(g)$, which is the maximum possible since $H_0(g)$ is an upper bound (see Eq.~\eqref{eq:MI-Inequalities}).  The reason for this result is that in this case the contingency table itself uniquely defines~$g$, so neglecting it puts the mutual information in error by an amount equal to the complete information cost of the ground truth. If we include the the information cost of the table, on the other hand, as we do in the reduced mutual information, this erroneous behavior disappears.  Assuming for simplicity that the ground-truth groups are of equal size, the optimal concentration parameters are $\alpha_{g|c} = 0$ and $\alpha_{g} = \infty$, and the reduced mutual information of Eq.~\eqref{eq:I-Approx} becomes
\begin{align}
I(c;g) &= I_0(c;g) - \log\left[\frac{n!}{\prod_{r} n_r!} (1/q_g)^n\right] - n \log q_g \nonumber\\
&= 0.
\end{align}
which is the correct answer.

Outside this special case, we observe from Eq.~\eqref{eq:I-Approx} that in the limit where the numbers of groups~$q_g = q_c = q$ are held fixed and~$n \rightarrow \infty$, the traditional mutual information grows as
\begin{align}
    I_0(c; g) \sim \Ord(n\log q )
\end{align}
while the correction term grows as
\begin{align}
    I(c;g) - I_0(c;g) \sim \Ord(q^2 \log n ).
\end{align}
Thus, in this limit the subleading term is relevant when $n/\log n \lesssim q^2/\log q$, a condition that holds in many practical contexts, where the number of communities can grow as $\sqrt{n}$ or faster.

\end{document}